\documentclass[sigconf]{acmart}

\usepackage{booktabs}
\usepackage{multirow}

\AtBeginDocument{%
  }

\setcopyright{acmlicensed}
\copyrightyear{2018}
\acmYear{2018}
\acmDOI{XXXXXXX.XXXXXXX}
\acmConference[Conference acronym 'XX]{Make sure to enter the correct
  conference title from your rights confirmation email}{June 03--05,
  2018}{Woodstock, NY}
\acmISBN{978-1-4503-XXXX-X/2018/06}

\begin{document}

\title{SepLRM: A Foundation Model Paradigm for Recommendation}
% \title{Separate LRM: A Foundation Model Paradigm for Recommendation}


\settopmatter{authorsperrow=1}

\author{%
Di Chen$^{*}$,
Xijun Xiao$^{*}$,
Qin Ren$^{*}$,
Weijiang Lai$^{*}$,
Zhijian Pan$^{*}$,
Zhentao Pan$^{*}$,
Pengfei Nie,\\
Cheng Yuan,
Zixuan Tan,
Xuyang Liu,
Tao Wang,
Yunwen Huang,
Hao Fang,\\
Yichao Wang,
Hao Zhang,
Siyuan Wu,
Hao Chen,
Han Diao,
Mingyu Yang,\\
Yongkang Feng,
Beihong Jin\textsuperscript{\S},
Jinfeng Yang,
Deping Xie,
Peng Xu,\\
Zheng Chai$^{\dagger}$,
Yuchao Zheng$^{\dagger}$%
}

\affiliation{%
  \institution{%
    ByteDance;
    \textsuperscript{\S} Institute of Software,
    Chinese Academy of Sciences%
  }
  \country{%
    \{chendi.666, xiaoxijun, renqin.97, laiweijiang,
    panzhijian, panzhentao, niepengfei.797,
    yuancheng.spoon, tanzixuan, liuxuyang.plt,
    wangtao.0911, huangyunwen.eleanor,
    fanghao.nice, wangyichao.0402, zhanghao.5577,
    wusiyuan.03, chenhao.spike, diaohan.nb666,
    yangmingyu, fengyongkang.00, yangjinfeng.02,
    deping.xie, xupeng, \\chaizheng.cz,
    zhengyuchao.yc\}@bytedance.com, beihong@iscas.ac.cn%
  }
}

\thanks{%
  * Equal contributions.\\
  % \qquad
  $\dagger$ Corresponding authors.
  % \qquad
  % \textsuperscript{\S} Institute of Software,
  %Chinese Academy of Sciences.
}

\renewcommand{\shortauthors}{Di Chen et al.}

\begin{abstract}
Large recommendation models (LRMs) improve recommendation performance by scaling behavior sequence length, model capacity, and training compute. However, the return on investment (ROI) of continued scaling deteriorates as marginal gains diminish and long-sequence encoding costs rise, especially when this computation is repeatedly executed by distinct models for diverse recommendation stages and business scenarios.
To address this issue, we propose SepLRM, a recommendation foundation model (RFM) paradigm that identifies reusable sequence encoding and reorganizes multiple end-to-end (E2E) LRMs into one RFM and multiple independently optimized vertical models (VMs).
The RFM adopts an encoder-decoder architecture and exports its final-layer encoder state as the shared key-value (KV) cache that is independent of downstream task-offered candidates. Unlike candidate-specific logits or aggregated user embeddings, the KV cache preserves fine-grained behavioral information. Separation at this native KV interface also allows each VM to construct candidate-conditioned queries while retaining the same query-KV interaction as its E2E counterpart, enabling high-fidelity capability transfer.
To ensure both transfer effectiveness and system efficiency, we co-design cache compression, the training and serving workflows, and LG-Store, a distributed storage system for KV caches.
Experiments on three large-scale datasets show that SepLRM achieves state-of-the-art offline performance with a transfer rate of approximately 100\%. It also demonstrates strong transferability across model iterations, recommendation stages, and business scenarios. Deployment in two large-scale industrial scenarios substantially improves the ROI of sequence scaling and delivers consistent gains in online A/B tests at ByteDance.
\end{abstract}





\maketitle

\section{Introduction}
In recommendation, advertising, and search systems, user behavior sequences are crucial for modeling user interests. Early sequential recommendation methods primarily model the relevance between historical behaviors and candidates. With advances in Transformers and inspiration from scaling laws~\cite{gpt4}, recent methods increasingly use self-attention to capture higher-order dependencies among behaviors and achieve continued performance gains by scaling sequence length, model capacity, and training compute. The scaled models are referred to as large recommendation models (LRMs).

Existing studies show the performance of LRMs improves sublinearly with FLOPs following a power law~\cite{scaling_law, scaling_law2}, whereas training and serving costs grow near-linearly or even superlinearly with computational scale. Consequently, continued scaling yields diminishing marginal gains and may eventually drive return on investment (ROI) below an acceptable business threshold. In industrial systems, this issue is amplified as multiple LRMs serving different recommendation stages and business scenarios repeatedly encode overlapping user behavior histories, causing platform-wide costs to grow with the number of LRMs. Overcoming this ROI constraint is therefore critical to further LRM scaling~\cite{roi1,roi2}.

%Existing methods reduce sequence modeling costs through 
Currently, methods to reduce sequence modeling costs include 
behavior sampling and retrieval, efficient attention, and operator optimization~\cite{longer,sparsectr,flashattention}. However, each E2E model still performs long-sequence encoding independently, preventing its cost from being amortized across models. Intermediate representation reuse within an LRM partially alleviates this issue by sharing computation across candidates, samples, or requests~\cite{ug-sep, lemur}, but remains confined to individual models. More recently, several studies have begun treating an LRM as a shared knowledge source for multiple independent models~\cite{rec-distill,llatte}, marking a transition from single-model scaling to multi-model capability reuse, and forming different recommendation foundation model (RFM) paradigms. However, existing RFMs typically rely on coarse-grained or candidate-conditioned interfaces that downstream models consume as supervision or fixed features, limiting intact 
%faithful 
transfer of the gains acquired through scaling.

To overcome these limitations, we propose SepLRM (\textbf{Sep}arate \textbf{L}arge \textbf{R}ecommendation \textbf{M}odels), an RFM paradigm that reorganizes multiple E2E LRMs on the same industrial platform into one shared RFM and multiple independently optimized vertical models (VMs). The RFM continually advances general sequence modeling through longer sequences, larger model capacity, and improved sequence modeling methods, while each VM independently evolves its candidate-conditioned information selection, feature interaction, and prediction modules. This separation allows general and scenario-specific modeling to evolve in parallel, with each RFM upgrade benefiting multiple VMs across recommendation stages and business scenarios.

%To overcome these limitations, we propose SepLRM, a RFM paradigm that separates multiple E2E LRMs into one shared RFM and multiple independently optimized downstream vertical models (VMs). The RFM adopts an encoder-decoder architecture: its encoder captures higher-order behavioral dependencies and exports a candidate-agnostic KV cache, while each VM constructs queries from its own candidate and request context to select relevant information through cross-attention. As a continually evolving shared sequence model, the RFM can scale to longer sequences, deeper architectures, and more advanced sequence-modeling methods, while each VM independently evolves its candidate-conditioned information selection, feature interaction, and prediction modules, allowing general sequence modeling and scenario-specific modeling to advance in parallel. Through the unified KV interface, each RFM improvement can benefit multiple VMs across recommendation stages and business scenarios, transforming single-model scaling gains into platform-wide benefits.

%SepLRM adopts an encoder-decoder architecture for capability-preserving model separation. In the RFM, 
In SepLRM, the RFM adopts an encoder-decoder architecture for capability-preserving model separation, where
the encoder performs general sequence modeling through self-attention, capturing higher-order behavioral dependencies and exporting the candidate-agnostic key-value (KV) cache. In each VM, the decoder constructs queries from its own candidate and request context and selects relevant information through cross-attention. Unlike decoder-only architectures that jointly model behaviors and candidates, separation at this native KV boundary preserves the query-KV interactions of the E2E LRMs, reducing architectural mismatch and capability transfer loss.

%In SepLRM, the encoder-decoder architecture is key to high-fidelity model separation. Unlike decoder-only architectures that jointly encode user behaviors and candidates from the outset, it decouples candidate-agnostic sequence encoding from candidate-conditioned information selection, preventing candidate information from entering the shared representations prematurely. Each VM therefore preserves the same query-KV interaction as its E2E counterpart, reducing architectural mismatch and transfer loss.

%From the FRM perspective, the RFM serves as a shared sequence model that continually evolves by scaling to longer behavior sequences, deeper architectures, and more advanced sequence-modeling methods. It exposes these capabilities to multiple independent VMs through a unified KV interface. Each VM evolves its candidate-conditioned information selection, feature interaction, and prediction modules for a specific scenario, allowing general sequence modeling and scenario-specific modeling to advance in parallel. Consequently, each RFM improvement benefits multiple downstream models, transforming single-model scaling gains into platform-wide gains.

As the knowledge interface between the RFM and VMs, the KV cache preserves finer-grained behavioral information than logits or aggregated user embeddings while remaining candidate-agnostic and downstream-queryable. Each VM can use its own queries to select and compose candidate-relevant information, achieving an approximately \(100\%\) transfer rate (TR) under lossless caching. To improve practical ROI, we further develop composable cache optimizations, including general compression and the user-level KV cache. By reducing per-cache size and cross-sample redundancy, these strategies provide a flexible trade-off between transfer effectiveness and storage cost while supporting knowledge reuse across recommendation stages and business scenarios.

To operationalize SepLRM, we co-design its training and serving workflows with the storage infrastructure for the KV caches. During training, an asynchronous pipeline overlaps cache I/O with model computation to hide cache-access latency. During serving, KV cache preparation runs in parallel with candidate retrieval and pre-ranking, moving most sequence-encoding overhead off the critical ranking procedure. We further develop LG-Store, which provides high-throughput, cost-efficient persistence and batch accesses for sharing KV caches across models, training jobs, and online services.

Our contributions are summarized as follows:
\begin{itemize}
\item We propose SepLRM, an RFM paradigm that separates E2E LRMs into one shared RFM and multiple independent VMs, reusing long-sequence encoding across models, stages, and scenarios to improve platform-wide scaling ROI.

\item We present the KV cache, which preserves fine-grained, candidate-agnostic user behavioral information and remains downstream-queryable. Together with native encoder-de\-cod\-er separation, it preserves E2E query-KV interactions and achieves a TR of approximately \(100\%\).

\item We co-design an asynchronous cache pipeline for training, KV cache preparation for serving, and LG-Store, a distributed store for shared KV caches, to reduce the storage, communication, and critical-path overhead of cache reuse and enhance the throughput of cache accesses.

% \item We co-design training and serving workflows with LG-Store, a distributed store for KV caches, to reduce the storage, communication, and critical-path overhead of cache reuse.

%\item Extensive offline experiments and online A/B tests show that SepLRM achieves state-of-the-art performance, enables transfer across models, stages, and scenarios, and substantially improves the ROI of sequence scaling.

\item Extensive experimental results
%experiments 
show that SepLRM achieves state-of-the-art performance and broad transferability. Online, it improves GIP ADSS/ADVV by 1.961\%/2.107\% and E-commerce GMV by 0.35\%-4.81\%, while achieving \(8.62\times/2.62\times\) the ROI of matched E2E LRMs, respectively.
\end{itemize}

\section{Related Work}
Research on scalable recommendation models has evolved from scaling individual LRMs to reusing intermediate representations within a model, and ultimately to sharing learned capabilities across independent models through the RFM paradigm. 
%The following subsections trace this progression.
\subsection{Large Recommendation Models}
%User behavior sequence modeling is fundamental to recommender systems. Early studies primarily captured user preferences through candidate-aware interest aggregation. DIN attends to historical behaviors relevant to the current candidate, DIEN further models the temporal evolution of user interests, and BST employs Transformers to capture higher-order behavioral dependencies \cite{din,dien,bst}. As user histories continue to grow, research has progressively shifted from short-term interest modeling toward long-sequence and even lifelong behavior modeling.

User behavior sequence modeling is fundamental to recommender systems. Early methods capture candidate-conditioned interests through target-attention, interest evolution, or Transformers~\cite{din,dien,bst}. As user histories grow, research has shifted toward long-sequence and lifelong behavior modeling.

To model longer behavior sequences, methods such as SIM and TWIN first retrieve candidate-relevant subsequences and then model the selected behaviors \cite{sim,eta,twin,twinv2,ubr4ctr,diffumin}.
%Although this two-stage paradigm reduces online computation, it limits modeling of the complete user-interest space.
Other methods retain full sequences while reducing computation through approximate or sparse attention, as exemplified by SDIM and SparseCTR \cite{sdim,sparsectr}.

Inspired by scaling laws in large language models, recent studies build LRMs by scaling model capacity, sequence length, and training computation. LONGER and SUAN enlarge sequence-modeling capacity \cite{longer,suan,sparsectr}; Wukong and RankMixer scale higher-order feature interaction \cite{wukong,rankmixer,rankmixerv2}; and HyFormer and OneTrans jointly model sequential and non-sequential features \cite{hyformer,onetrans}. HSTU and OneRec further explore unified modeling across recommendation tasks or stages \cite{hstu,onerec,onerecv2,openonerec}.

While these studies demonstrate consistent gains from scaling, the resulting computation remains independently repeated across models, increasingly constraining the ROI of further LRM scaling.
\subsection{LRMs with Representation Reuse}
%\subsection{Recommendation Models with Representation Reuse}
%\subsection{Reusable Intermediate Representations}
To reduce repeated computation, several studies cache and reuse intermediate representations. 
%MIMN maintains long-term user interests in a memory network, avoiding repeated processing of complete behavior histories \cite{mimn},
%LONGER precomputes and caches sequence-side keys and values for reuse across candidates \cite{longer}, 
MARM caches intermediate states of complex sequence modules to trade storage for lower online FLOPs \cite{marm}. Similarly, UG-Sep separates user behavior modeling from candidate-conditioned information flow in dense interaction networks, enabling candidate-agnostic representations to be reused across samples and candidates \cite{ug-sep}. SequenceO1 compresses ultra-long histories into a fixed number of candidate-agnostic sketch tokens and processes them through candidate-conditioned cross-attention, enabling reuse across candidates, training samples, and consecutive online requests \cite{sequenceo1}.

%These studies demonstrate the value of intermediate representation reuse for scaling recommendation models. However, such reuse is largely confined to computations within a single model and does not provide a general representation interface shared by independent downstream models. Consequently, expensive computation can only be amortized within a limited scope, restricting cross-model and cross-scenario reuse and the resulting ROI gains.
%These studies demonstrate the value of intermediate representation reuse, but largely confine it within a single model rather than exposing a shared interface to independent downstream models. This limited reuse scope restricts cross-model, cross-stage cross-scenario cost amortization and the resulting ROI gains.
However, these methods confine representation reuse within a single model, limiting cost amortization and ROI gains.

\begin{figure*}[tb]
  \includegraphics[width=0.9\textwidth]{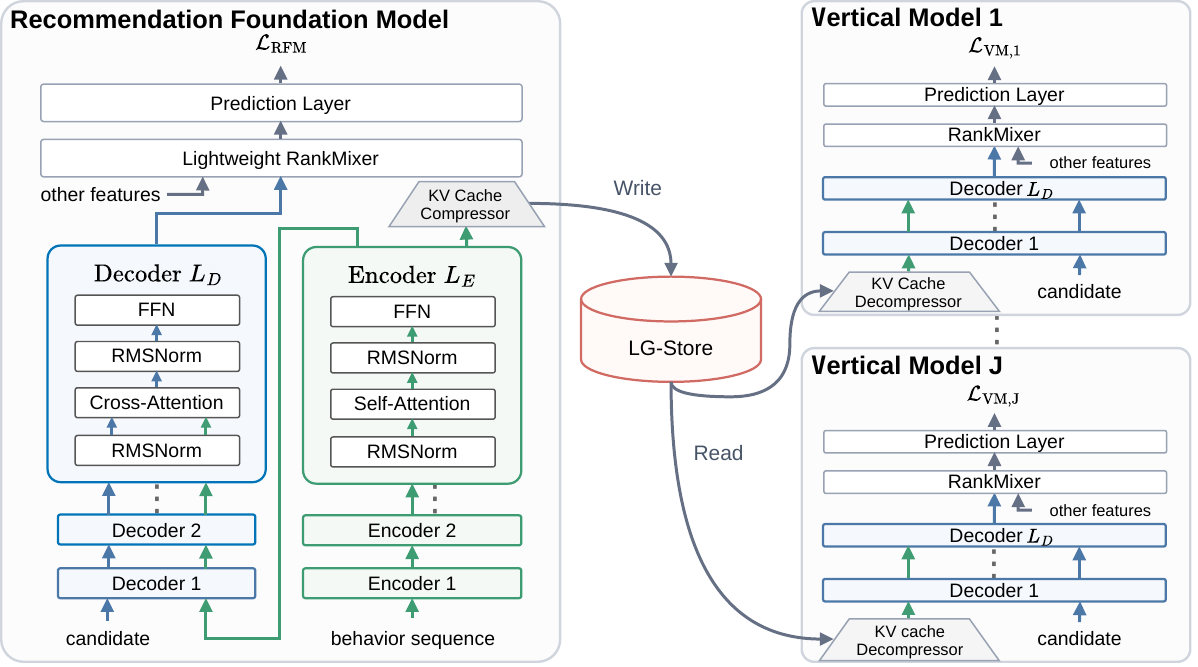}
  \caption{Architecture of SepLRM.}
  \label{fig:model}
\end{figure*}
\subsection{Recommendation Foundation Models}
To move beyond the single-model scaling boundary of task-specific LRMs, the RFM paradigm turns a high-capacity LRM into a shared knowledge source and transfers knowledge to multiple downstream models. Existing studies explore different transfer mechanisms. Rec-Distill caches logits produced by a large teacher as supervision for multiple students across ranking stages~\cite{rec-distill}. LoopFM further extracts and compresses multi-layer representations from historical samples into sequence features \cite{loopfm}. LLaTTE and Foundation-Expert transfer knowledge through compressed user representations and candidate-conditioned representations, respectively \cite{llatte,realizing}.

Existing methods rely on coarse-grained or candidate-conditioned interfaces that downstream models consume as supervision or fixed features, limiting generality and downstream flexibility. In contrast, SepLRM provides the fine-grained, candidate-agnostic, and downstream-queryable KV cache that 
compresses a user’s historical behaviors and 
preserves the native query-KV interactions, enabling high-fidelity transfer across models, stages, and scenarios.

% Despite this progress, existing approaches remain limited. Logits and aggregated representations are coarse-grained and may create information bottlenecks, while candidate-conditioned intermediate representations have limited generality. Moreover, downstream models typically consume these outputs as supervision or fixed features, preventing them from flexibly selecting and composing fine-grained behavioral information for their own candidates and tasks.

% In contrast, SepLRM exposes a fine-grained, candidate-agnostic KV cache as a downstream-queryable sequence memory. Each VM can select and compose candidate-relevant behavioral information through cross-attention. By preserving the native decoder query-KV interaction across model separation, SepLRM provides a structural foundation for high-fidelity transfer across models, stages, and scenarios.

\section{Methodology}
\subsection{Overview}
Given a user \(u\), we denote the historical behavior sequence by \(\mathcal B_u=\{b_1,b_2,\ldots,b_n\}\), where each behavior contains features such as item ID and behavior timestamp. In addition to sequential behaviors, the user is associated with non-sequential features \(\mathcal U_u=\{u_1,u_2, \ldots,u_k\}\), such as age, gender. For a candidate \(c\), \(\mathcal C_c=\{c_1,c_2,\\ \ldots,c_m\}\) represents the candidate and contextual features, including item ID and request time. Here, \(n\), \(k\), and \(m\) denote the number of historical behaviors, non-sequential user features, and the total number of candidate and contextual features, respectively.

A training sample is represented as \(s=(\mathcal B_u,\mathcal U_u,\mathcal C_c,\mathbf y)\), where \(\mathbf y\) collects the supervision labels for the prediction tasks. Given \(s\), a conventional E2E LRM produces task predictions as follows:
\begin{equation}
    \hat{\mathbf y}_{\mathrm{E2E}}=P_{\mathrm{E2E}}\left(\mathcal B_u,\mathcal U_u,\mathcal C_c;\Theta_{\mathrm{E2E}}\right),
\end{equation}
where \(\Theta_{\mathrm{E2E}}\) denotes its model parameters. As sequence length and model capacity scale, repeated long-sequence encoding across multiple E2E LRMs rapidly increases platform-wide computation costs and constrains the ROI of sequence scaling.

As illustrated in Figure~\(\ref{fig:model}\), we propose SepLRM to restructure \(J\) independent LRMs on the same platform into one shared RFM and \(J\) VMs. The RFM performs its own prediction tasks while exporting its final-layer encoder states as a reusable KV cache \(\mathcal{KV}_s\), formally:
\begin{equation}
    \left( \hat{\mathbf y}_{\mathrm{RFM}}, \mathcal{KV}_s \right) = P_{\mathrm{RFM}} \left( \mathcal B_u,\mathcal U_u,\mathcal C_c; \Theta_{\mathrm{RFM}} \right). 
\end{equation}
%Although the RFM uses candidate features for its own predictions, \(\mathcal{KV}_s\) is generated solely by its candidate-agnostic encoder. 
The \(j\)-th VM consumes \(\mathcal{KV}_s\) with its own features without repeating long-sequence encoding, and its prediction is given by:
\begin{equation}
    \hat{\mathbf y}_{\mathrm{VM},j} = P_{\mathrm{VM},j} \left( \mathcal{KV}_s, \mathcal U_u, \mathcal C_c; \Theta_{\mathrm{VM},j} \right).
\end{equation}
%The RFM and VMs are independently optimized using their respective data and supervision objectives. 
For each model, the weighted multi-task objective is defined as:
\begin{equation}
    \mathcal L = \sum_t \lambda_t\, \ell_t \left( y^{(t)},\hat y^{(t)} \right),
\end{equation}
where \(y^{(t)}\) and \(\hat y^{(t)}\) are the \(t\)-th components of \(\mathbf y\) and \(\hat{\mathbf y}\), respectively, and \(\ell_t\) and \(\lambda_t\) denote the corresponding loss function and task weight.

The RFM and VMs are independently optimized on their own data and objectives, where the former specializes in sequence modeling, with configurable sequence lengths and encoder capacities, and the latter focus on their candidate-conditioned information selection, feature interaction, and prediction modules. 

% Through this separation, the RFM concentrates on scalable long-sequence modeling, while each VM retains its own task objectives and model architecture and reuses the shared sequence representations through a unified KV interface.
% Through this separation, SepLRM transforms sequence scaling from an isolated capability of individual models into a reusable system-level capability. The RFM serves as a shared sequence model that continuously scales to learn transferable long-sequence representations, while each Downstream LRM retains its own task objectives and model architecture and reuses the RFM’s sequence-modeling capability through a unified KV interface.
\subsection{Recommendation Foundation Model}
The RFM consists of input embedding, an encoder-decoder sequence modeling backbone, and a task prediction module, with the encoder exporting its final states as a reusable KV cache.
% The RFM adopts an encoder-decoder architecture. Its encoder captures higher-order behavioral dependencies through self-attention and exports the final states as a candidate-agnostic KV cache, while its decoder queries the cache using the candidate and request context through cross-attention. 
%This native separation boundary preserves the query-KV interaction and reduces transfer loss.
%The RFM adopts an encoder-decoder architecture that separates candidate-agnostic behavior encoding from candidate-conditioned information selection. The encoder captures high-order behavioral dependencies through multi-layer self-attention and exports the final-layer representations as a KV cache, while its decoder constructs queries from the candidate and request context to select relevant information through cross-attention and produce predictions. The KV interface forms a native separation boundary that preserves query-KV interaction across model separation, thereby reducing transfer loss.

\subsubsection{Input Embedding}
Each behavior is embedded from its item ID, behavior type, and other features. The resulting behavior sequence is represented by \(\mathbf H^0=\operatorname{Emb}(\mathcal B_u)\in\mathbb R^{n\times d}\), where \(d\) is the embedding dimension. The candidate and contextual features form \(q\) query tokens, denoted by \(\mathbf C^0=\operatorname{Emb}_{\mathrm C}(\mathcal C_c)\in\mathbb R^{q\times d}\). Similarly, the non-sequential user features are represented by \(\mathbf O=\operatorname{Emb}_{\mathrm U}(\mathcal U_u)\in\mathbb R^{k\times d}\). The encoder takes \(\mathbf H^0\) as input, the decoder uses \(\mathbf C^0\) as its initial queries, and \(\mathbf O\) is incorporated into feature interaction and prediction.

\subsubsection{Encoder Block}
The encoder stacks \(L_E\) blocks to transform behavior embeddings into contextualized representations, capturing general, candidate-agnostic temporal and higher-order dependencies from each behavior’s visible history.
%The encoder stacks \(L_E\) blocks to transform the initial behavior embeddings into contextualized sequence representations, allowing each behavior to incorporate information from its visible history and capture temporal and higher-order dependencies.

Each encoder block adopts a pre-norm architecture with multi-head self-attention, a SwiGLU feed-forward network, and residual connections. In the \(l\)-th block, the input \(\mathbf H^{l-1}\) is first normalized as \(\mathbf X^l=\operatorname{RMSNorm}(\mathbf H^{l-1})\) and then processed by \(R\)-head self-attention, where each head has dimension \(d_r=d/R\). The self-attention computation is given by:
\begin{equation}
    \mathbf Q_r^l=\mathbf X^l\mathbf W_{Q,r}^l,\qquad
\mathbf K_r^l=\mathbf X^l\mathbf W_{K,r}^l,\qquad
\mathbf V_r^l=\mathbf X^l\mathbf W_{V,r}^l,
\end{equation}
\begin{equation}
    \mathbf A_{r}^{l} = \operatorname{Softmax} \left( \frac{ \mathbf Q_{r}^{l} \left(\mathbf K_{r}^{l}\right)^\top }{ \sqrt{d_r} } + \mathbf M \right) \mathbf V_{r}^{l}, 
\end{equation}
\begin{equation}
    \widetilde{\mathbf H}^l = \mathbf H^{l-1} + \operatorname{Concat} \left( \mathbf A_{1}^l, \ldots, \mathbf A_{R}^l \right) \mathbf W_O^l, 
\end{equation}
where \(\mathbf W_{Q,r}^l,\mathbf W_{K,r}^l,\mathbf W_{V,r}^l\) and \(\mathbf W_O^l\) are learnable projections, and \(\mathbf M\) is the causal mask. We further normalize the attention output as \(\widetilde{\mathbf X}^{l}=\operatorname{RMSNorm}(\widetilde{\mathbf H}^l)\). The subsequent SwiGLU feed-forward transformation and residual update are formulated as~\cite{swiglu}:
\begin{equation}
\mathbf H^l = \widetilde{\mathbf H}^l + \left[ \operatorname{SiLU} \left( \widetilde{\mathbf X}^l\mathbf W_1^l \right) \odot \left( \widetilde{\mathbf X}^l\mathbf W_2^l \right) \right]\mathbf W_3^l, 
\end{equation}
where \(\mathbf W_1^l,\mathbf W_2^l\in\mathbb R^{d\times d_{\mathrm{ff}}}\) and \(\mathbf W_3^l\in\mathbb R^{d_{\mathrm{ff}}\times d}\) are learnable projection matrices, \(d_{\mathrm{ff}}\) is the intermediate dimension, and \(\odot\) denotes element-wise multiplication. 

%After \(L_E\) encoder blocks, the final hidden states \(\mathbf H^{L_E}\in\mathbb R^{n\times d}\) incorporate sequential context while retaining behavior-level granularity. They are consumed by the RFM decoder and compressed into the shared representation for VMs:
After \(L_E\) encoder blocks, the contextualized behavior states \(\mathbf H^{L_E}\in\mathbb R^{n\times d}\) are consumed by the RFM decoder and compressed into the shared representation for VMs:
\begin{equation}
    \mathcal{KV}_s=\operatorname{Comp}_{\phi}\!\left(\mathbf H^{L_E}\right)\in\mathbb R^{n\times d_c},
\end{equation}
where \(\operatorname{Comp}_{\phi}(\cdot)\) denotes a lightweight down-projection followed by quantization, and \(d_c \leq d\) is the KV-cache embedding dimension.
Interestingly, our design shares a architectural
principle with the causal encoder-decoder architecture of
DeepSeek-V4.1-Flash~\cite{deepseekv41}: using a shared representation
as the source of decoder keys and values. Specifically, we use the
shared \(\mathcal{KV}_s\) as a compact KV cache rather than storing
layer-specific key and value tensors. Each VM decoder independently
projects it into the keys and values required by its cross-attention layers.

% Our design shares principles with the architecture of DeepSeek-V4.1-Flash~\cite{deepseekv41}. Specifically, we use the shared \(\mathcal{KV}_s\) as a compact KV cache rather than storing layer-specific key and value tensors. Each VM decoder independently projects it into the keys and values required by its cross-attention layers.

%Inspired by the architecture of DeepSeek-V4.1-Flash~\cite{deepseekv41} and the joint KV compression in multi-head latent attention~\cite{deepseekv2}, we use the shared \(\mathcal{KV}_s\) as the KV cache, rather than layer-specific key and value tensors. Each VM decoder independently projects it into the keys and values required by its cross-attention layers.

%Following joint KV compression in multi-head latent attention~\cite{deepseekv2}, we use the shared \(\mathcal{KV}_s\) as the KV cache, rather than layer-specific key and value tensors. Each VM decoder independently projects it into the keys and values required by its cross-attention layers.

%Throughout this paper, KV cache refers to the shared latent representation \(\mathcal{KV}_s\), rather than layer-specific projected keys and values. Each decoder independently projects it into the keys and values required by its cross-attention layers. Similar to MLA’s joint KV compression \cite{deepseekv2}, this design supports low-rank projection, quantization, and other compression strategies to reduce storage and transmission costs.

\subsubsection{Decoder Block}
The RFM stacks \(L_D\) decoder blocks, in which candidate and request context features serve as queries in cross-attention over \(\mathbf H^{L_E}\). This candidate-conditioned interaction progressively selects relevant behavioral information and produces candidate-specific user interest representations.

Each decoder block adopts a similar pre-norm architecture as the encoder block, with self-attention replaced by cross-attention. In the \(l\)-th block, the candidate representation is normalized as \(\mathbf X_{\mathrm C}^{l}=\operatorname{RMSNorm}(\mathbf C^{l-1})\). The query, key, and value projections for the \(r\)-th head are given by:
\begin{equation}
    \mathbf Q_r^l = \mathbf X_{\mathrm C}^{l}\mathbf W_{Q,r}^{l}, \qquad \mathbf K_r^l = \mathbf H^{L_E}\mathbf W_{K,r}^{l}, \qquad \mathbf V_r^l = \mathbf H^{L_E}\mathbf W_{V,r}^{l}.
\end{equation}
Here, \(\mathbf W_{Q,r}^{l},\mathbf W_{K,r}^{l},\mathbf W_{V,r}^{l}\in\mathbb R^{d\times d_r}\) are learnable projection matrices. The queries are progressively updated across layers, while each layer attends to the same encoder states through layer-specific projections. The remaining attention, residual, and SwiGLU operations follow Section 3.2.2. The final representation \(\mathbf C^{L_D}\) is passed to the task prediction module.

\subsubsection{Task Prediction}
Although the RFM primarily learns reusable long-sequence representations, we retain a lightweight feature interactions module and task-specific prediction heads to provide effective supervision, narrow the modeling gap between the RFM and VMs, and improve the transferability of the KV cache.

After \(L_D\) decoder blocks, the candidate-conditioned representation \(\mathbf C^{L_D}\) is concatenated with the non-sequential feature representation \(\mathbf O\in\mathbb R^{k\times d}\), yielding \(\mathbf Z^0=\mathbf C^{L_D}\Vert\mathbf O\in\mathbb R^{(q+k)\times d}\), where \(\Vert\) denotes concatenation. We process these tokens using \(L_{RM}\) RankMixer blocks~\cite{rankmixer}, which combine token mixing and nonlinear feature transformations to capture higher-order feature interactions. After pooling, the prediction for task \(t\) is computed as:
% We concatenate it with the non-sequential feature representations \(\mathbf O\in\mathbb R^{k\times d}\) to form \(\mathbf Z^0=\mathbf C^{L_D}\Vert\mathbf O\in\mathbb R^{(q+k)\times d}\), where \(\Vert\) denotes concatenation. The resulting tokens are processed by \(L_R\) RankMixer blocks \cite{rankmixer} to model higher-order feature interactions. The \(l\)-th block is defined as:
% \begin{equation}
%     \widetilde{\mathbf Z}^{l} = \operatorname{LN}_{1}^{l} \left( \mathbf Z^{l-1} + \operatorname{TokenMix}^{l} \left(\mathbf Z^{l-1}\right) \right)
% \end{equation}
% \begin{equation}
%     \mathbf Z^{l} = \operatorname{LN}_{2}^{l} \left( \widetilde{\mathbf Z}^{l} + \operatorname{PFFN}^{l} \left(\widetilde{\mathbf Z}^{l}\right) \right)
% \end{equation}
% where \(\operatorname{LN}(\cdot)\) denotes layer normalization, \(\operatorname{TokenMix}(\cdot)\) enables cross-feature interaction by reorganizing feature subspaces across tokens, and \(\operatorname{PFFN}(\cdot)\) applies nonlinear transformations to individual tokens. After pooling the interacted token representations, the prediction for task \(t\) is given by:
\begin{equation}
    \hat y_{\mathrm{RFM}}^{(t)} = g_{\mathrm{RFM}}^{(t)} \left( \operatorname{Pool} \left(\mathbf Z^{L_{RM}}\right) \right),
\end{equation}
where \(g_{\mathrm{RFM}}^{(t)}(\cdot)\) denotes the prediction head for task \(t\). Here, we adopt RankMixer for feature interaction, while SepLRM remains compatible with alternative feature interaction architectures.

\subsection{Vertical Model}
Each VM consumes the candidate-agnostic KV cache exported by the RFM, avoiding repeated long-sequence encoding. The cached representation is first reconstructed as:
\begin{equation}
    \widehat{\mathbf H}_{s} = \operatorname{Decomp}_{\psi} \left( \mathcal{KV}_s \right) \in\mathbb R^{n\times d},
\end{equation}
where \(\operatorname{Decomp}_{\psi}(\cdot)\) denotes dequantization followed by a lightweight up-projection to the decoder representation space. Each VM then constructs queries from its candidate and context features and applies the cross-attention decoder described in Section 3.2.3 to select relevant information from \(\widehat{\mathbf H}_s\). Its decoder independently projects the KV cache into layer-specific keys and values.

The decoder output is combined with the VM’s non-sequential features for task prediction. Different VMs may use their own decoders, feature interaction modules, and prediction heads, enabling the KV cache to be reused across models, recommendation stages, and business scenarios.

\begin{figure}[tb]
  \includegraphics[width=\linewidth]{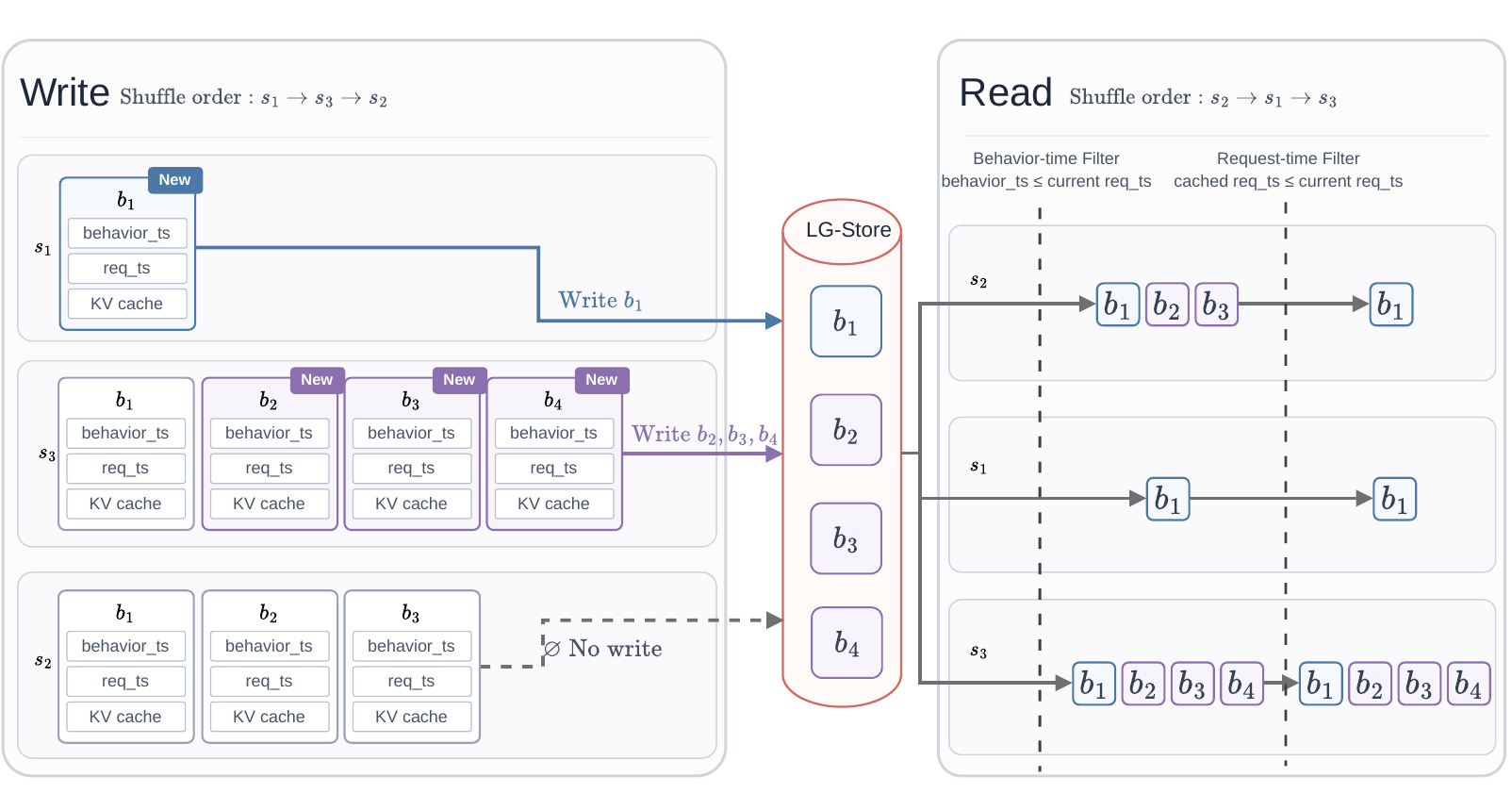}
  \caption{Illustration of the User-level KV Cache.}
  \label{fig:append}
\end{figure}
\subsection{KV Caches}
Fine-grained KV caches enable high-fidelity transfer but incur substantial storage and communication costs. SepLRM mitigates these costs through two complementary designs: general cache compression and user-level KV cache compression.

\subsubsection{General Cache Compression}
Under the sample-level KV cache, each sample $s$ with $\texttt{insid}$ as its identifier is stored as \(\texttt{insid}\rightarrow\mathcal{KV}_s\). SepLRM stores only final encoder states and further compresses them through low-rank projection from \(d\) to \(d_c\) dimensions and quantization from \(b\) to \(b_c\) bits.

%SepLRM further reduces total cache volume through staged access. Instead of caching the entire training period, the RFM writes KV cache entries only during later training, and each VM consumes them only during its corresponding later stage.
On the other hand, SepLRM further reduces total cache volume through staged access. Specifically, the RFM writes KV cache entries only in the late training stage (referred to as the cache coverage period), rather than throughout training. Thus, only samples within the cache coverage period need to be stored. Moreover, each VM consumes them only during its corresponding later training stage. In production, this strategy achieves the same performance as full-period cache access, while the coverage period can be configured according to data scale, training schedule, and convergence behavior of the scenario. For \(N\) training samples, of which \(N_c\) are cached, the overall compression ratio is \(\Gamma_{\mathrm{cache}}=N_cd_cb_c/(Ndb) \times 100\%\).

%General cache compression reduces the storage and communication costs of the KV cache. SepLRM first exports only the representations derived from the final encoder states, avoiding intermediate states from other encoder layers. Under sample-level cache, each sample is stored as \(\texttt{insid}\rightarrow\mathcal{KV}_s\).

%Each cache can be compressed by reducing its dimension from \(d\) to \(d_c\) through low-rank projection and its precision from \(b\) to \(b_c\) bits through quantization, according to the storage budget and target TR. SepLRM further reduces cache volume through staged access: the RFM writes KV caches only during later training, and each VM consumes them only during its corresponding later stage. In production, this strategy achieves the same performance as full-period cache writing and consumption. The cache coverage period is configured for each scenario based on its data scale, training schedule, and convergence behavior. For \(N\) training samples, of which \(N_c\) are cached, the overall compression ratio over full-period uncompressed cache is \(\Gamma_{\mathrm{cache}}=Ndb/(N_c d_c b_c)\).

\subsubsection{User-level Cache Compression}
%Although general compression reduces individual cache size, the sample-level KV cache still duplicates overlapping behavior windows and is tied to aligned samples. 
%Furthermore, we design a user-level KV cache keyed by user identifier \(\texttt{uid}\), which removes cross-sample redundancy and enables reuse across unaligned data streams, recommendation stages, and business scenarios: \(\texttt{uid}\rightarrow[(\mathcal{KV}_{u}, \texttt{req\_ts},\texttt{behavior\_ts})]\), where \(\texttt{behavior\_ts}\) and \(\texttt{req\_ts}\) denote the timestamps of user behavior and request, respectively.

Furthermore, we design a user-level KV cache: \(\texttt{uid}\rightarrow[(\mathcal{KV}_{u}, \texttt{req\_ts},\texttt{behavior\_ts})]\), where \(\texttt{uid}\) denotes the user identifier, \(\texttt{behavior\_ts}\) and \(\texttt{req\_ts}\) denote the timestamps of user behavior and request, respectively. 
This cache removes cross-sample redundancy and enables reuse across unaligned data streams, recommendation stages, and business scenarios.

As shown in Figure~\ref{fig:append}, behavior\_ts is used to append only new behavior representations. Because training shuffling may reorder samples, cache reads apply behavior-time and request-time filters to exclude future behaviors and caches generated by later requests, respectively. Strict filtering may leave recent behaviors unavailable; for example, \(s_2\) recovers \(b_1\) but not \(b_2\) and \(b_3\). To mitigate this gap between the cached and recent behaviors, each VM also employs a lightweight encoder to encode recent behaviors and combines them with the cached history in its decoder.

%Each VM mitigates this gap by encoding recent behaviors with a lightweight encoder and combining them with the cached history in its decoder.

Thus, sample-level cache offers higher TR at the expense of increased cross-sample redundancy, whereas user-level cache reduces storage and supports reuse across data streams, stages, and scenarios with a slight compromise in transfer effectiveness. The final choice depends on the storage budget, desired TR, and reuse scope.

\begin{figure}[tb]
  \includegraphics[width=0.9\linewidth]{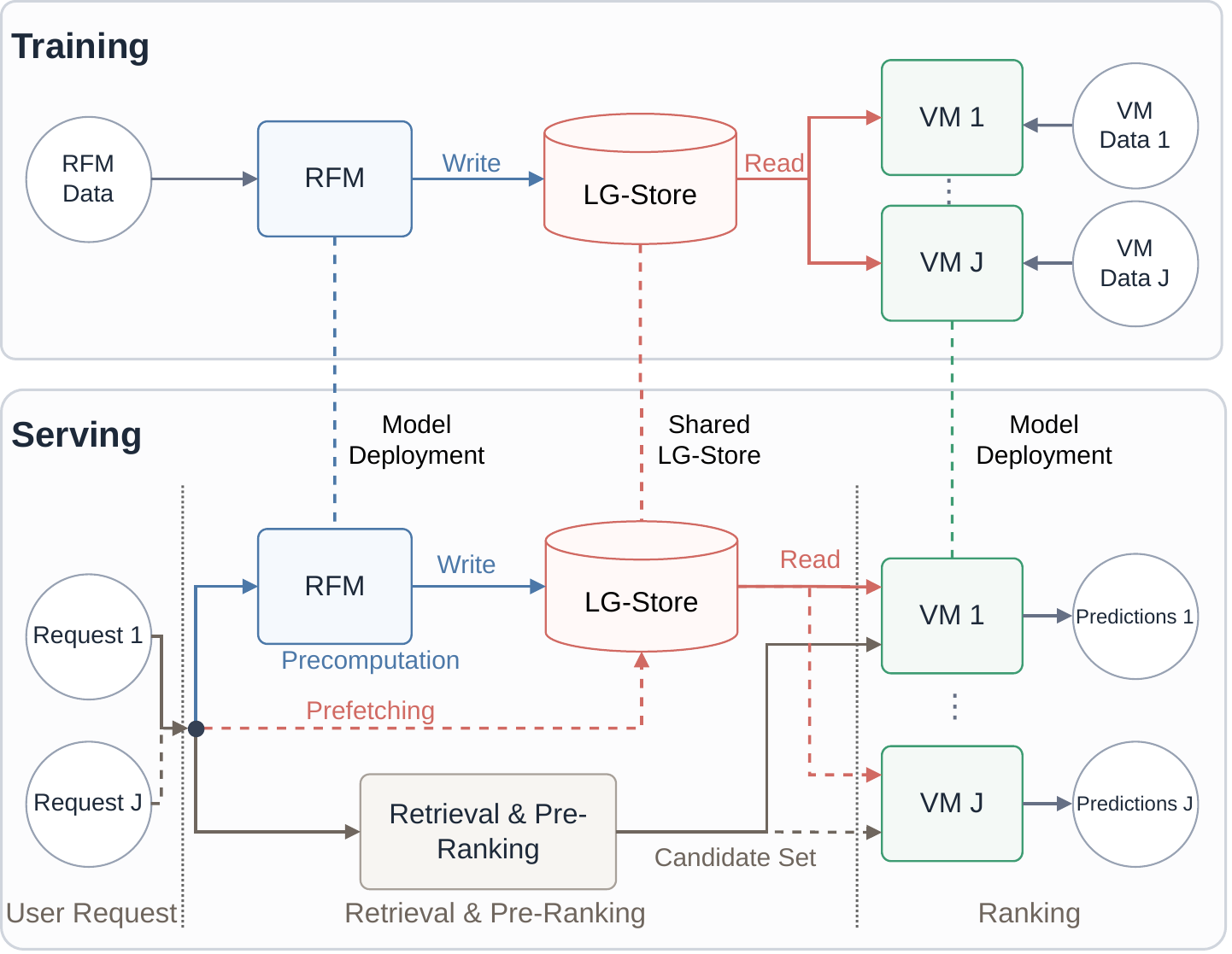}
  \caption{Training and serving workflows of SepLRM.}
  \label{fig:workflows}
\end{figure}
\section{System}
%We operationalize SepLRM with coordinated training and serving workflows and LG-Store. We further analyze its computational cost, storage complexity, and ROI.

\subsection{Training}
As shown in Figure~\ref{fig:workflows}, the RFM and VMs are trained independently, but both require access to the KV caches. As the shared sequence model, the RFM prioritizes its capacity for sequence modeling: stronger feature interaction may improve its own predictions without benefiting VMs, whereas better sequence modeling directly enhances the shared representations. We therefore retain only lightweight feature interaction and prediction heads in the RFM to provide effective supervision.

To reduce the I/O and communication overhead for KV cache access, an asynchronous pipeline is designed to overlap cache operations for adjacent mini-batches with current model computation. Since thread-level asynchrony still competes for memory bandwidth, and CPU time and cache, we also isolate cache access and data processing in separate processes, leaving the training process to submit requests and receive KV tensors. Besides, we limit request concurrency to prevent backlogs. In brief, this pipeline improves training throughput and stability. 

%To reduce KV cache I/O and communication overhead, an asynchronous pipeline overlaps cache operations for adjacent mini-batches with current model computation. Since thread-level asynchrony still competes for CPU scheduling, cache, and memory bandwidth, we isolate cache access and data processing in separate processes, leaving the training process to submit requests and receive KV tensors. We also limit request concurrency to prevent backlogs, improving training throughput and stability.

\subsection{Serving}
KV cache preparation is launched before downstream ranking and overlaps it with candidate generation and pre-ranking. It supports two modes: precomputation, which invokes the RFM upon request arrival to generate an up-to-date cache, and prefetching, which loads an existing KV cache from LG-Store with freshness determined by its update frequency. The unified KV interface is provided to allow different VMs to choose the mode based on the freshness requirement and computation cost.

Once candidates enter downstream ranking, each VM constructs queries from the candidates and request context and selects relevant information from the prepared KV caches through cross-attention. This procedure moves most sequence encoding, cache access, and communication overhead off the critical ranking path.

\subsection{LG-Store}
%Model separation introduces large-scale KV cache storage, access, and synchronization demands that conventional storage systems cannot serve cost effectively. We thus design LG-Store (Lagrange-Store), a distributed storage system to share RFM representations across VMs.

%LG-Store (Lagrange-Store), a distributed storage system, is proposed for the efficient storage and cross-VM sharing of large-scale KV caches.

LG-Store (Lagrange-Store) is a high-throughput, cost-efficient distributed storage system designed for large-scale KV cache storage and cross-VM sharing.

LG-Store combines hash partitioning, an in-memory index, and append-only files for efficient point and batched lookups, avoiding the multilevel lookups and amplification of LSM trees. It separates the solid-state drive (SSD) cache from persistent storage, allowing read throughput and storage capacity to scale independently. For high fan-in reads, LG-Store aggregates keys by their destination nodes and leverages callback-driven refill to control in-flight requests and bytes, thus reducing RPC overhead, network congestion, and tail latency. For cross-cluster synchronization, it partitions KV caches by user\_id and aggregates and filters the latest valid entries before transmission, avoiding redundant or stale traffic.

In production workloads, LG-Store reduces average batch read latency by 64\%, lowers cross-cluster traffic to 0.6\%-1\% of direct synchronization, and costs only 43.5\% as much as a conventional disaggregated KV system under the same throughput, durability, and tail-latency requirements.

\subsection{Efficiency Analysis}
Due to limited space, only computational overhead of long-sequence modeling, cache storage overhead and ROI analysis are presented.
%We analyze SepLRM in terms of computation, storage, and ROI. 
%Consider \(J\) comparable E2E LRMs, which SepLRM reorganizes into one shared RFM and \(J\) VMs.
%For simplicity, we consider only long-sequence modeling and cache overhead.

\subsubsection{Computational Complexity Analysis}
Ignoring constant factors, the encoder and decoder complexities are \(\Phi_{\mathrm{Enc}}=\mathcal O(L_E(n^2d+nd^2+ndd_{\mathrm{ff}}))\) and \(\Phi_{\mathrm{Dec}}=\mathcal O(L_D(qnd+(n+q)d^2+qdd_{\mathrm{ff}}))\). Since \(q\ll n\), the encoder dominates the cost of sequence scaling. The \(J\) E2E LRMs require approximately \(J(\Phi_{\mathrm{Enc}}+\Phi_{\mathrm{Dec}})\), In SepLRM, the encoder is executed once by the shared RFM, while the decoder is executed once by the RFM and once by each of the \(J\) VMs. The resulting computation is therefore \(\Phi_{\mathrm{Enc}}+(J+1)\Phi_{\mathrm{Dec}}+\Phi_{\mathrm{KV}}\). Here, \(\Phi_{\mathrm{KV}}\) denotes cache compression and processing overhead, which is substantially smaller than \(\Phi_{\mathrm{Enc}}\). SepLRM saves approximately \(J-1\Phi_{\mathrm{Enc}}\).

During serving, the latency in E2E LRM is \(\tau_{\mathrm{Cand}}+\tau_{\mathrm{Enc}}+\tau_{\mathrm{Down}}\), where the three terms represent the latencies for candidate retrieval and pre-ranking, sequence encoding, and downstream prediction latency, respectively. By overlapping KV cache preparation with candidate retrieval and pre-ranking, SepLRM reduces the latency to \(\max(\tau_{\mathrm{Cand}},\tau_{\mathrm{Prep}})+\tau_{\mathrm{Down}}\). When \(\tau_{\mathrm{Prep}}\leq\tau_{\mathrm{Cand}}\), it decreases by approximately \(\tau_{\mathrm{Enc}}\).

\subsubsection{Storage Analysis}
SepLRM stores model parameters and the additional KV cache. By sharing one encoder, it eliminates \(J-1\) duplicate encoder parameter sets compared with conventional E2E LRMs. For the sample-level KV cache, the storage complexity is \(\mathcal O(\Gamma_{\mathrm{cache}}Nndb)\), where \(\Gamma_{\mathrm{cache}}\) denotes the overall compression ratio. The user-level KV cache further removes duplicated behavior representations across samples of the same user.

\subsubsection{ROI Analysis}
Beyond theoretical complexity, we compare SepLRM and conventional E2E LRMs using \(\operatorname{ROI}=\Delta\mathcal G/\Delta\mathcal K\), where \(\Delta\mathcal G\) is the incremental gain over the baseline whose cost is negligible compared to any LRM and \(\Delta\mathcal K\) is the corresponding training, serving, and storage cost.

%Compared with independently scaling the \(J\) E2E LRMs, the incremental costs 
Starting from this baseline, the incremental costs for forming \(J\) E2E LRMs and corresponding models of SepLRM
are approximately \(\Delta\mathcal K_{\mathrm{E2E}}=J(\mathcal K_{\mathrm{Enc}}+\mathcal K_{\mathrm{Dec}})\) and \(\Delta\mathcal K_{\mathrm{SepLRM}}=\mathcal K_{\mathrm{Enc}}+(J+1)\mathcal K_{\mathrm{Dec}}+\mathcal K_{\mathrm{KV}}\), where \(\mathcal K_{\mathrm{KV}}\) covers KV cache storage, transfer, and access. We further define the gain retention and cost ratios as \(\eta_{\mathrm{gain}}=\Delta\mathcal G_{\mathrm{SepLRM}}/\Delta\mathcal G_{\mathrm{E2E}}\) and \(\eta_{\mathrm{cost}}=\Delta\mathcal K_{\mathrm{SepLRM}}/\Delta\mathcal K_{\mathrm{E2E}}\), respectively.
With approximately \(100\%\) TR, \(\eta_{\mathrm{gain}}\approx1\), yielding \(\operatorname{ROI}_{\mathrm{SepLRM}}/\operatorname{ROI}_{\mathrm{E2E}}  \approx\Delta\mathcal K_{\mathrm{E2E}}/\Delta\mathcal K_{\mathrm{SepLRM}}\). This advantage grows with sequence modeling FLOPs and the number of VMs as more encoding cost is amortized.

%With approximately \(100\%\) TR, SepLRM nearly preserves the effectiveness gain of its E2E counterpart, yielding \(\eta_{\mathrm{gain}}\approx1\). Their relative ROI can therefore be expressed as \(\operatorname{ROI}_{\mathrm{SepLRM}}/\operatorname{ROI}_{\mathrm{E2E}}=\eta_{\mathrm{gain}}/\eta_{\mathrm{cost}}\approx\Delta\mathcal K_{\mathrm{E2E}}/\Delta\mathcal K_{\mathrm{SepLRM}}\). As sequence length, encoder capacity, or the number of VMs increases, SepLRM amortizes more repeated encoding cost, further widening its ROI advantage.

\begin{table*}[tbp]
\caption{
Performance comparison. AUC values are averaged over the seven-day test period. The best and second-best results in each row are marked in bold and underlined, respectively.
%denoted by bold and underlined text, respectively.
NoSeq serves as the base model for computing $\Delta\mathrm{AUC}$.
}
%\vspace{-6pt}
\label{tab:performance_comparison}
\centering
\begin{tabular}{cc|cccccc|cc}
\toprule
Dataset & Metric
& NoSeq & SIM & TWIN & HSTU & LONGER & LLaTTE
& E2E LRM & SepLRM \\
\midrule

\multirow[c]{2}{*}{Douyin E-commerce}
& AUC
& 0.84131
& 0.84805
& 0.84808
& 0.84833
& 0.84870
& \underline{0.84909}
& \textbf{0.85010}
& \textbf{0.85010} \\
& $\Delta\mathrm{AUC}$ (\%)
& +0.000
& +0.674
& +0.677
& +0.702
& +0.739
& \underline{+0.778}
& \textbf{+0.879}
& \textbf{+0.879} \\

\midrule

\multirow[c]{2}{*}{Hongguo}
& AUC
& 0.84162
& 0.84484
& 0.84558
& 0.84677
& 0.84719
& \underline{0.84740}
& \textbf{0.84798}
& \textbf{0.84798} \\
& $\Delta\mathrm{AUC}$ (\%)
& +0.000
& +0.322
& +0.396
& +0.515
& +0.557
& \underline{+0.578}
& \textbf{+0.636}
& \textbf{+0.636} \\

\midrule

\multirow[c]{2}{*}{KuaiRand-27K}
& AUC
& 0.64640
& 0.67357
& 0.67449
& 0.67841
& 0.68034
& 0.68142
& \underline{0.68307}
& \textbf{0.68308} \\
& $\Delta\mathrm{AUC}$ (\%)
& +0.000
& +2.717
& +2.809
& +3.201
& +3.394
& +3.502
& \underline{+3.667}
& \textbf{+3.668} \\

\bottomrule
\end{tabular}
\end{table*}

\section{Experiments}
%Since SepLRM aims to transfer and reuse the sequence scaling gains of the RFM, w
%We evaluate SepLRM by VMs primary-task AUC and transfer rate. Sections 5.2-5.4 examine downstream effectiveness, transfer across stages and scenarios, and scaling ROI, respectively. Section 5.5 then analyzes how RFM architecture and KV cache design affect VM performance.
%We evaluate SepLRM using the primary-task AUC and transfer rate of its VMs. We first examine downstream effectiveness, transfer across stages and scenarios, and scaling ROI, and then analyze the effects of RFM architecture and KV cache design.
\subsection{Experimental Settings}
\subsubsection{Datasets}
We evaluate SepLRM on two large-scale industrial datasets and KuaiRand-27K \cite{kuairand}.
%, each containing hundreds of millions of interactions.

The industrial datasets are collected from Douyin e-commerce and Hongguo short-drama scenarios. We use data from three consecutive months for training and the following seven days for testing. The maximum sequence lengths are 10,000 and 5,000, respectively.

KuaiRand-27K contains 322M interactions between 27K users and 32M videos. We chronologically split the samples into training, validation, and test sets at a ratio of 8:1:1. The maximum behavior sequence length is set to 5,000.

\subsubsection{Competitors and Implementation Details}
We instantiate SepLRM as an RFM with a 6-layer encoder-decoder and a VM containing a 6-layer decoder for comparison with other baselines. The baselines are drawn from two categories. The first includes the E2E sequence modeling baselines SIM~\cite{sim}, TWIN~\cite{twin}, HSTU~\cite{hstu}, and LONGER~\cite{longer}. The second is RFM paradigm baselines, e.g., LLaTTE~\cite{llatte}, which transfers compressed user embeddings to downstream models. 
In addition, we introduce two reference baselines for quantifying sequence scaling and TR: E2E LRM, which integrates the same long-sequence architecture and capacity as SepLRM but is trained E2E, and NoSeq, its counterpart with the long-sequence modeling component removed from E2E LRM.

For a fair comparison, we rebuild the E2E models by preserving their original sequence modeling components while aligning all other components with SepLRM VM. Moreover, SepLRM and all baselines use the same model configurations, optimizer, and training hyperparameters. For example, HSTU, LONGER, and LLaTTE, are uniformly configured with six layers.

%We compare SepLRM with five representative baselines from two research lines. The E2E sequence modeling baselines include SIM, TWIN, HSTU, and LONGER \cite{sim,twin,hstu,longer}. We further include LLaTTE \cite{llatte} as an RFM paradigm baseline, which transfers compressed user embeddings to downstream models. In addition, we introduce two reference configurations for quantifying sequence scaling and TR: E2E LRM, which integrates the same long-sequence architecture and capacity as SepLRM but is trained E2E, and NoSeq, its counterpart with the long-sequence modeling component removed.

%For a fair comparison, all methods differ only in their sequence modeling components and otherwise use the same model configurations, optimizer, and training hyperparameters. On the two industrial datasets, the hidden dimension is set to 512, and a four-layer RankMixer is used for feature interaction. Stackable sequence backbones, including HSTU, LONGER, and LLaTTE, are uniformly configured with six layers, while both E2E LRM and SepLRM adopt six-layer encoders and decoders (6E6D). The industrial and KuaiRand-27K experiments were conducted with 128 and 1 A100-equivalent accelerators, respectively.

\subsubsection{Evaluation Metrics}
For offline evaluation, we report AUC as the main metric. For RFMs, including LLaTTE and SepLRM, we report the AUC of their downstream models. Notably, in our industrial settings, an absolute AUC improvement of 0.0001 is considered practically meaningful~\cite{hyformer,longer}. To quantify how much of the E2E long-sequence gain an RFM preserves in its downstream model, we define the transfer rate (TR) as: 
\begin{equation}
    \operatorname{TR} = \frac{ \operatorname{AUC}_{\mathrm{VM}} - \operatorname{AUC}_{\mathrm{NoSeq}} }{ \operatorname{AUC}_{\mathrm{E2E }} - \operatorname{AUC}_{\mathrm{NoSeq}} } \times 100\%
\end{equation}
where \(\operatorname{AUC}_{\mathrm{VM}}\), \(\operatorname{AUC}_{\mathrm{NoSeq}}\), and \(\operatorname{AUC}_{\mathrm{E2E}}\) denote the AUCs of the VM, its counterpart without long-sequence modeling, and the corresponding E2E LRM, respectively.

For online evaluation, we report relative improvements over the corresponding production baselines in advertiser score (ADSS) and advertiser value (ADVV) for advertising, and orders and gross merchandise volume (GMV) for e-commerce.
%For online evaluation, we conduct A/B tests and report relative improvements in core business metrics, including \(\mathrm{CTR/CVR/GMV}]\) for Douyin E-commerce and \(\mathrm{watch\ time/valid\ play}]\) for Hongguo short-drama.

\subsection{Comparison with Competitors}
Table~\ref{tab:performance_comparison} compares the primary-task AUC of SepLRM with competing methods across the three datasets. All long-sequence methods outperform NoSeq, confirming the value of modeling long-term behavioral histories. Moreover, LRM-based methods, including HSTU, LONGER, E2E LRM, and the two RFMs, consistently outperform conventional long-sequence methods, demonstrating the additional benefits of scaling sequence length and model capacity.

Among RFMs, LLaTTE remains inferior to SepLRM. It compresses long-term behaviors into a coarse-grained, fixed user embedding, while its downstream model queries only its own short sequence, resulting in substantial information loss. SepLRM instead retains the complete behavior sequence in fine-grained KV caches, allowing downstream queries to access the full history through native query–KV interactions. Consequently, SepLRM achieves the best performance among RFMs.

%Among RFMs, LLaTTE remains inferior to SepLRM, suggesting that compressing an entire behavior sequence into a fixed user embedding introduces an information bottleneck and limits candidate-conditioned retrieval. SepLRM instead adopts a fine-grained, candidate-agnostic, and downstream-queryable KV cache as the knowledge interface, while separating the encoder and decoder at the native KV boundary. This combination preserves behavior-level information and the same query-KV interaction as the scaled E2E LRM, enabling SepLRM to achieve the strongest downstream performance among RFMs.
%and a TR of approximately 100\% across all datasets.

\begin{table}[t]
\centering
\caption{Transfer performance of SepLRM KV cache.}
%\vspace{-6pt}
\label{tab:transfer_performance}
\begin{tabular}{ccc}
\toprule
\textbf{Transfer Setting}
& \textbf{Method}
& \textbf{$\Delta$AUC ($\times 10^{-4}$)} \\
\midrule

\multirow{2}{*}{\shortstack{Cross-stage:\\E-commerce pre-ranking}}
& Base
& -- \\
& Base + KV cache
& +70 \\

\midrule

\multirow{2}{*}{\shortstack{Cross-scenario:\\Mall ranking}}
& Base
& -- \\
& Base + KV cache
& +10 \\

\bottomrule
\end{tabular}
\end{table}
\subsection{Transfer across Stages and Scenarios}
To evaluate whether the long-sequence modeling capability of our RFM can extend to diverse settings, we reuse the user-level KV caches generated by the RFM, which is trained on  e-commerce ranking data, in two additional settings: e-commerce pre-ranking and mall ranking. 
Reusing KV caches is highly convenient, since user-level indexing bridges unaligned data streams, while the candidate-agnostic representation allows each VM to query the shared caches with its own candidates and request context in its decoder.
%To evaluate whether our RFM’s long-sequence modeling capability can extend beyond its primary setting, we reuse the user-level KV cache generated by the E-commerce RFM in two additional settings: E-commerce pre-ranking and mall ranking. User-level indexing bridges unaligned data streams, while the candidate-agnostic representation allows each VM to query the shared cache with its own candidates and request context through an added decoder. 

As shown in Table~\ref{tab:transfer_performance}, SepLRM yields substantial gains in pre-ranking, where full long-sequence modeling is often impractical because of the large candidate set and strict latency budget. By reusing the KV caches and introducing only a lightweight decoder, SepLRM enables long-sequence modeling without repeating expensive encoder computation.

%Full long-sequence modeling is often impractical in pre-ranking because of its large candidate set and strict latency budget. By reusing the KV cache and introducing only a lightweight decoder, SepLRM enables long-sequence modeling without repeating expensive encoder computation, yielding substantial gains in this resource-constrained stage.

We further reuse the KV caches for mall ranking. Despite differences in feature distributions and supervision objectives from the e-commerce ranking, the observed improvement demonstrates that the shared sequence representation remains effective across scenarios. Together, these results show that heterogeneous VMs can consume the same KV caches, enabling one RFM to support multiple recommendation stages and scenarios.

\begin{figure}[tb]
  \includegraphics[width=0.75\linewidth]{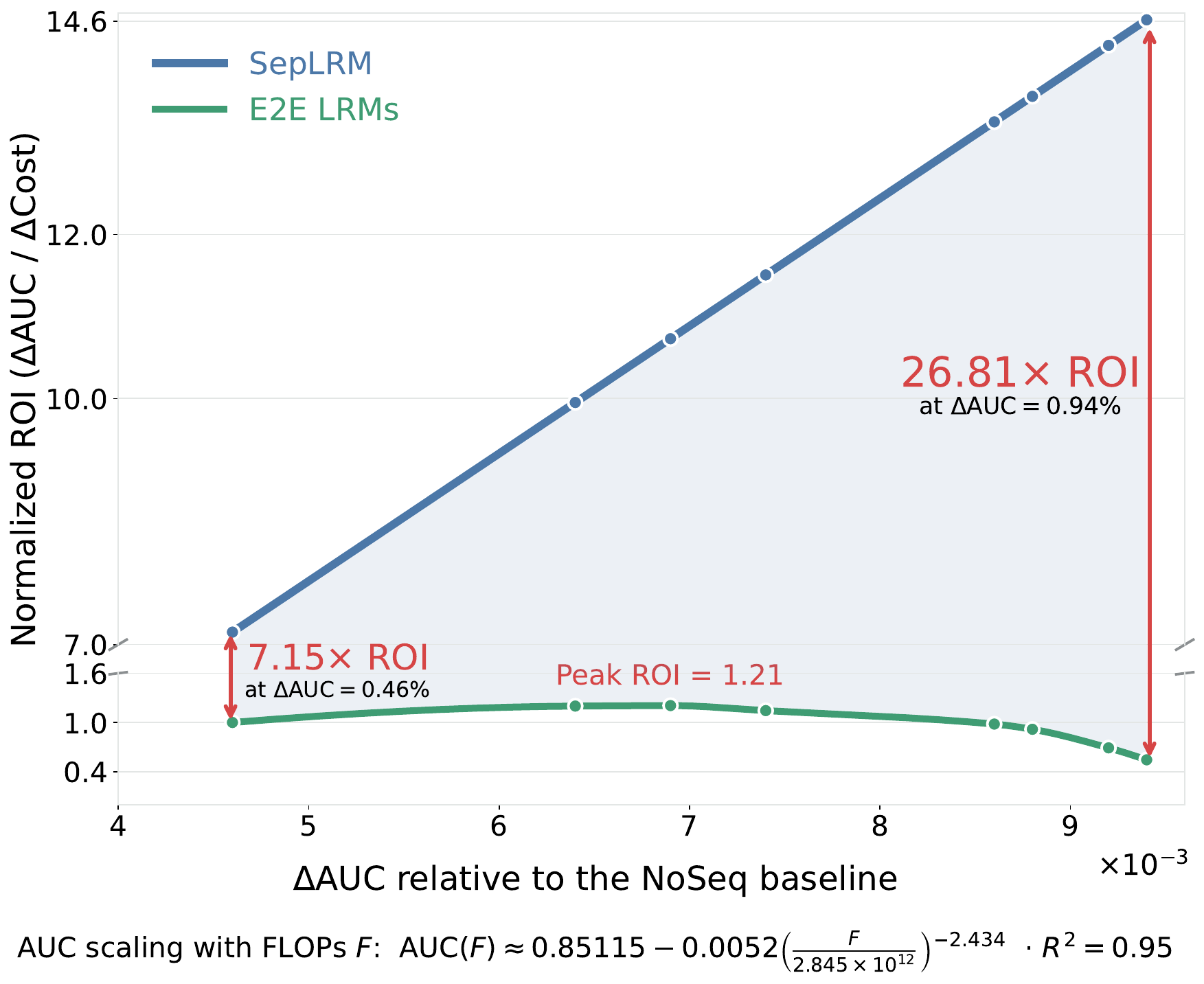}
  \caption{Sequence Scaling ROI of SepLRM and E2E LRMs.}
  \label{fig:roi}
\end{figure}

\subsection{Sequence Scaling and ROI Analysis}
To study how long-sequence recommendation scales and whether SepLRM improves its ROI, we construct matched E2E LRM and SepLRM versions on the e-commerce dataset by scaling the sequence length from 0.2K to 10K and the encoder-decoder depth from 2E2D to 8E8D. Using NoSeq as the baseline, we measure training FLOPs, AUC, and resource costs to fit the scaling function and calculate offline ROI.

As training FLOPs \(F\) increase to 7.0\(\times 10^{12}\), the AUC gain over NoSeq reaches \(9.4\times10^{-3}\). The measured results are well described by the power-law function \(\operatorname{AUC}(F)\approx0.85115-0.0052(F/(2.845\times10^{12}))^{-2.434}\), with \(R^2=0.95\). This relationship confirms that scaling sequence length and model capacity steadily improves AUC, albeit with diminishing returns from additional computation.

%For each configuration \(m\), we define the incremental gain as \(\Delta\mathcal G_m=\operatorname{AUC}_m-\operatorname{AUC}_{\mathrm{NoSeq}}\), the incremental cost as \(\Delta\mathcal K_m=C_m-C_{\mathrm{NoSeq}}\), and the offline ROI as \(\operatorname{ROI}_m=\Delta\mathcal G_m/\Delta\mathcal K_m\). We consider \(J=30\) LRMs: E2E LRMs independently incur the full sequence-modeling cost, whereas SepLRM shares one RFM across its VMs. Training costs are based on measured accelerator-hours and unit prices, with SepLRM additionally including 30 days of KV cache storage. All ROI values are normalized to that of the smallest E2E LRM.

For each version with its specific configuration \(f\), we define the offline incremental gain and cost over NoSeq as \(\Delta\mathcal G_f=\operatorname{AUC}_f- \\ \operatorname{AUC}_{\mathrm{NoSeq}}\) and \(\Delta\mathcal K_f=\mathcal K_f-\mathcal K_{\mathrm{NoSeq}}\), respectively, with \(\operatorname{ROI}_f=\Delta\mathcal G_f/\Delta\mathcal K_f\). We evaluate 30 LRMs, a typical system scale in our industrial settings, where SepLRM shares one RFM across 30 VMs instead of independently scaling every E2E LRM. \(\mathcal K\) is calculated from measured accelerator hours and their unit price, with 30 days of KV cache storage additionally included for SepLRM.
%Costs \(\mathcal K\) are measured accelerator-hours multiplied by unit price, plus 30 days of KV cache storage for SepLRM. 
The ROI is normalized to that of the E2E LRM at the minimal configuration.
%ROI is normalized to the smallest E2E configuration.

As shown in Figure~\ref{fig:roi}, the ROI of E2E LRMs rises initially but soon peaks and declines as marginal AUC gains diminish while encoder costs grow. In contrast, the ROI of SepLRM steadily increases,
%through the use of cross-model encoder, 
widening its advantage from \(7.15\times\) to \(26.81\times\). This trend suggests that the ROI gap will widen further as sequence length and model capacity continue to scale.

%As shown in Figure~\ref{fig:roi}, E2E LRM ROI initially increases but soon peaks and declines increasingly rapidly as marginal AUC gains diminish while encoder costs continue to grow. In contrast, SepLRM steadily improves its ROI through cross-model encoder reuse, expanding its advantage from 7.15\(\times\) at the smallest scale to 26.81\(\times\) at the largest. This trend demonstrates that SepLRM increasingly amortizes scaling costs, making longer sequences and larger models more economically viable.

\begin{table}[t]
\centering
\caption{Performance of RFMs with different configurations.}
%{Performance comparison of different RFM configurations.}
%\vspace{-6pt}
\label{tab:RFM_configuration}
\setlength{\tabcolsep}{3pt}
\begin{tabular}{ccccc}
\toprule
\multicolumn{2}{c}{\textbf{RFM Configuration}}
& \multirow[c]{3}{*}{\textbf{RFM AUC}}
& \multirow[c]{3}{*}{\textbf{VM AUC}}
& \multirow[c]{3}{*}{\textbf{TR}} \\
\cmidrule(lr){1-2}
\multirow[c]{2}{*}{\textbf{Architecture}}
& \textbf{Feature}
& & & \\
& \textbf{Interaction}
& & & \\
\midrule
Encoder--Decoder & 4B   & 0.85010 & 0.85010 & 100\% \\
Encoder--Decoder & 700M & 0.84883 & 0.85010 & 100\% \\
Encoder--Decoder & None & 0.83213 & 0.84969 & 95\%  \\
Encoder-only     & 4B   & 0.85011 & 0.84918 & 89\%  \\
Decoder-only     & 4B   & 0.85017 & 0.84926 & 90\%  \\
\bottomrule
\end{tabular}
\end{table}
\subsection{Factors Affecting Transfer Rate}
\subsubsection{Effect of RFM Architecture}
To analyze how the RFM architecture affects TR, we vary its feature-interaction capacity and sequence-modeling architecture while keeping the VM and all other training settings fixed. The default RFM uses an encoder-decoder architecture with a 4B-parameter feature-interaction module.

As shown in Table~\ref{tab:RFM_configuration}, parameter reduction or even complete removal of the feature-interaction module in RFM lowers its own AUC but only marginally affects TR, which remains at \(95\%\) without the module. This suggests that transferable KV caches do not require large-scale feature interaction, so the parameter budget for sequence modeling should be prioritized in RFM.

Variants of the encoder-decoder architecture leaves RFM AUC nearly unchanged but substantially reduces TR. 
%In the encoder-decoder architecture, the candidates in the RFM and VM interacts only with the encoder's final-layer behavior representation. 
In encoder-only and decoder-only RFMs, the candidate interacts with the corresponding behavior representation at each layer. This layer-wise interaction cannot be reproduced by transferring only the final-layer cache. Transferring all layer-wise K/V states restores \(100\%\) TR in our validation, but substantially increases storage and communication costs. In contrast, the encoder-decoder architecture provides a natural transfer boundary between the encoder and decoder, achieving high TR by transferring only the encoder's final-layer behavior representation rather than the representations from every layer.

These results show that RFM AUC is not a reliable metric for KV cache quality, but VM AUC and TR are. Moreover, they further indicate that transfer effectiveness depends more on whether encoder-decoder separation preserves the native query-KV interaction, while additional capacity should be allocated preferentially to sequence modeling. 
%therefore be evaluated by VM AUC and TR.

\begin{table}[t]
\centering
\caption{Effects of KV cache design.}
%\vspace{-6pt}
%\caption{Effects of KV cache design on compression ratio \(\Gamma_{\mathrm{cache}}\) and transfer effectiveness.}
\label{tab:cache_design}
\setlength{\tabcolsep}{2pt}

\begin{tabular}{ccccc}
\toprule
\textbf{Cache Design}
& \textbf{Configuration}
& \textbf{\(\Gamma_{\mathrm{cache}}\)}
& \textbf{VM AUC}
& \textbf{TR} \\
\midrule

\multirow[c]{2}{*}{Quantization}
& INT8 & 50\% & 0.85013 & 100\% \\
& INT4 & 25\% & 0.85001 & 99\% \\

\midrule

\multirow[c]{2}{*}{Low-rank Projection}
& 256-dim & 50\% & 0.84974 & 96\% \\
& 128-dim & 25\% & 0.84952 & 93\% \\

\midrule

Cache Granularity
& User-level & 1.3\% & 0.84936 & 92\% \\

\bottomrule
\end{tabular}
\end{table}

\subsubsection{Effect of KV Cache Design}
To examine the trade-off between storage cost and transfer effectiveness, we vary the dimension, precision, and granularity of the KV cache, with all other settings fixed. The default configuration uses the sample-level KV cache with 512-dimensional BF16 representations.

As shown in Table~\ref{tab:cache_design}, SepLRM is robust to dimension reduction and low-bit quantization. These results reveal that quantization reduces storage with negligible transfer loss, while low-rank projection retains a high TR with modest degradation. Besides, the user-level KV cache further removes redundancy across samples, substantially reducing storage while preserving most of the sequence scaling gain. In SepLRM, these strategies can be combined according to storage, transfer, and reuse requirements.

\begin{table}[t]
\centering
\caption{Online A/B test results.} %on GIP Advertising and Douyin E-commerce.}
%\vspace{-6pt}
\label{tab:online_ab_test}
\begin{tabular}{lcc}
\toprule

\multicolumn{3}{l}{\textbf{(a) GIP Advertising}} \\
\addlinespace[3pt]
\textbf{Application}
& \textbf{ADSS}
& \textbf{ADVV} \\
\addlinespace[2pt]

Overall & +1.961\% & +2.107\% \\
Hongguo      & +2.240\% & +2.492\% \\
Toutiao      & +1.790\% & +1.470\% \\
Tomato Novel & +1.311\% & +1.500\% \\

\midrule

\multicolumn{3}{l}{\textbf{(b) Douyin E-commerce}} \\
\addlinespace[3pt]
\textbf{Content Format}
& \textbf{Order/U}
& \textbf{GMV/U} \\
\addlinespace[2pt]

Short Video    & +4.6306\% & +4.8077\% \\
Live Streaming & +0.3673\% & +0.3524\% \\

\bottomrule
\end{tabular}
\end{table}

\subsection{Online A/B tests}
We deploy SepLRM on ByteDance’s GIP (General Information Platform) Advertising and Douyin E-commerce, with sequence length and encoder-decoder depth configured as 10K/4E4D and 5K/6E6D, respectively. In each experiment, the baseline is the mature, highly optimized production model. The experiment on GIP Advertising uses 20\% of production traffic for nine days, while the experiment on Douyin E-commerce uses 30\% for 14 days; both cover hundreds of millions of users. We report relative gains over production controls using ADSS and ADVV for advertising, and Order/U and GMV/U for e-commerce. Table~\ref{tab:online_ab_test} summarizes the A/B test results.

On GIP Advertising, SepLRM consistently improves both ADSS and ADVV at the platform level and across representative applications, demonstrating robust gains across diverse content environments. On Douyin E-commerce, SepLRM improves both live-streaming and short-video commerce. Together, these experiments demonstrate that SepLRM delivers consistent gains across diverse large-scale production scenarios.

For ROI comparison, we additionally evaluate matched E2E LRMs at the same deployment scales. The ROIs of E2E LRMs and SepLRM are 26 and 224 on GIP Advertising, and 319 and 835 on Douyin E-commerce, respectively, corresponding to \(8.62\times\) and \(2.62\times\) those of the matched E2E LRMs.

\section{Conclusion}
We present SepLRM, an RFM paradigm that transforms sequence scaling from a model-specific capability into a reusable capability shared across distinct recommendation models. By separating E2E LRMs at the native encoder-decoder KV interface, SepLRM enables a shared RFM to export  fine-grained, candidate-agnostic KV caches that VMs can directly query, supporting high-fidelity reuse across models, stages, and scenarios. Moreover, combined with cache compression, training and serving workflows, and LG-Store, SepLRM preserves predictive performance while substantially improving scaling ROI.

\begin{acks}
We thank Song Wang, Chuanhua Xu, Jinghan Zhang, Kaiqi Tian, Hanqing Xie, Wei Jiang, Mizhipeng Zhang, Lei Zhang, Qingyuan Li, Lihan Ou, Yankai Wang, and Chuhan Zhao for their support of this work.
\end{acks}

\bibliographystyle{ACM-Reference-Format}
\bibliography{ref}
\appendix
\end{document}